\documentclass[sigconf]{acmart}

\AtBeginDocument{%
  \providecommand\BibTeX{{%
    \normalfont B\kern-0.5em{\scshape i\kern-0.25em b}\kern-0.8em\TeX}}}

\setcopyright{acmcopyright}
\copyrightyear{2018}
\acmYear{2018}
\acmDOI{XXXXXXX.XXXXXXX}

\acmConference[Conference acronym 'XX]{Make sure to enter the correct
  conference title from your rights confirmation emai}{June 03--05,
  2018}{Woodstock, NY}
\acmPrice{15.00}
\acmISBN{978-1-4503-XXXX-X/18/06}

\usepackage{enumitem}

\usepackage{xcolor}
\newif\ifcomment
\commenttrue
\commentfalse
\ifcomment
\newcommand{\shirin}[1]{{\bf \textcolor{red}{Shirin: #1}}}
\newcommand{\ana}[1]{{\bf \textcolor{blue}{Ana: #1}}}
\else
\newcommand{\shirin}[1]{}
\newcommand{\ana}[1]{}
\fi

\begin{document}

\title{Sadness, Anger, or Anxiety: Twitter Users’ Emotional Responses to Toxicity in Public Conversations}


\author{Ana Aleksandric}
\affiliation{%
  \institution{University of Texas at Arlington}
  \city{Arlington}
  \state{Texas}
  \country{United States}}

  \author{Hanani Pankaj}
\affiliation{%
  \institution{University of Texas at Arlington}
  \city{Arlington}
  \state{Texas}
  \country{United States}}

  \author{Gabriela Mustata Wilson}
\affiliation{%
  \institution{University of Texas at Arlington}
  \city{Arlington}
  \state{Texas}
  \country{United States}}

  \author{Shirin Nilizadeh}
\affiliation{%
  \institution{University of Texas at Arlington}
  \city{Arlington}
  \state{Texas}
  \country{United States}}

\renewcommand{\shortauthors}{Trovato and Tobin, et al.}

\begin{abstract}

Cyberbullying and online harassment have serious negative psychological and emotional consequences for the victims, such as decreased life satisfaction, suicidal ideation, self-harming behaviors, depression, anxiety, and others. 
Most of the prior works assessed people's emotional responses via questionnaires, while social media platforms contain data that could provide valuable insights into users' emotions in real online discussions.
Therefore, this data-driven study investigates the effect of toxicity on Twitter users' emotions and other factors associated with expressing anger, anxiety, and sadness in terms of account identifiability, activity, conversation structure, and conversation topic. To achieve this goal, we identified toxic replies in the large dataset consisting of 79,799 random Twitter conversations and obtained the emotions expressed in these conversations. Then, we performed propensity score matching 
and analyzed causal associations between toxicity and users’ emotions. In general, we found that users receiving toxic replies 
are more likely to express emotions of anger, sadness, and anxiety compared to users who did not receive toxic replies. 
 Finally, analysis results indicate that the conversation topic and users' account characteristics are likely to affect their emotional responses to toxicity. Our findings provide a better understanding of toxic replies' consequences on users' emotional states, which can potentially lead to developing personalized moderation methods that will help users emotionally cope with toxicity on social media.

\end{abstract}



\maketitle

\section{Introduction}

Previous literature highlighted some of the consequences of online harassment, cyberbullying, and trolling on the psychological well-being of victims~\cite{hinduja2007offline, kowalski2014bullying, giumetti2013rude}, who expressed psychological distress, decreased life satisfaction, suicidal ideation~\cite{giumetti2022cyberbullying}, self-harming behaviors, depression and anxiety~\cite{EYUBOGLU2021113730,ijerph17010045,Alhajji}. Cyber victimization might also have emotional consequences by triggering anger and sadness in users~\cite{elipe2015perceived}, whereas negatively coping with anger might lead to further cyberbullying behavior~\cite{den2016can}. Therefore, understanding the emotional needs of users in toxic conversations would be the first step in helping users with emotion regulation to prevent future cyberbullying attempts. 

Most of these studies, however, asked their participants to describe their emotional responses via questionnaire~\cite{ortega2012emotional} or they performed experiments to simulate cyberbullying exposure~\cite{alhujailli2020affective}. 
However, social media data gives an opportunity for observation of how users express emotions in real online discussions~\cite{stieglitz2013emotions,duncombe2019politics}, providing a great starting point for large-scale data-driven studies to deeply investigate users' emotions.
In this work, we examine whether the emotions of social media users change with the presence of toxicity, as well as the effect that amount of toxicity has on the emotions of users who received toxic replies. For example, we test whether users who received toxic replies to their tweets show a higher level of anger compared to users who did not receive any toxic replies. 

In statistical analysis, we assessed factors that could potentially affect users' emotions, such as conversation structure, emotions expressed before receiving toxicity, and conversation topics. In addition, models consider the effects of users' characteristics, i.e., online visibility, identifiability, and activity level. The dataset used in the study consists of a random sample of 79.8k Twitter conversations from August 14th to September 28th, 2021.
Moreover, each conversation was represented as a reply tree where two tweets are connected if one is a reply to another. In order to detect toxic replies, we compared the labels obtained from Google's Perspective API~\cite{jigsaw2021perspective} and OpenAI API~\cite{openaidoc} to manually labeled samples. After finding that Perspective API performs better, we used it to classify the replies as toxic or not toxic in the whole dataset. Then, we used LIWC-22, a software for analyzing word use~\cite{boyd2022development}, to identify the emotions of each post. Afterward, we employed propensity score matching to find users with similar account characteristics. As a result, we obtained a balanced dataset containing two groups of similar users: \emph{treatment} group and \emph{control} group, where each pair of users has a similar propensity (probability) of receiving toxic replies, isolating the effect of toxicity. Finally, we performed appropriate statistical tests to find a causal association between receiving toxicity and users' emotions. Note that causal associations describe relationships between variables where a causal link is suggested, but it does not make a definitive causal inference. 
The following hypotheses were formulated to get a better understanding of the \emph{causal associations} among our dependent and independent variables:
\begin{itemize}[leftmargin=*]
    \item[] \textbf{H1}: Users receiving toxic replies are more likely to express anxiety compared to users who did not receive any toxic replies.
     \item[] \textbf{H2}: Users receiving toxic replies are more likely to express anger compared to users who did not receive any toxic replies.
      \item[] \textbf{H3}: Users receiving toxic replies are more likely to express sadness compared to users who did not receive any toxic replies.
    \item[] \textbf{H4}: A larger amount of toxicity will likely increase users' anxiety. 
    \item[] \textbf{H5}:  A larger amount of toxicity will likely increase users' anger. 
    \item[] \textbf{H6}:  A larger amount of toxicity will likely increase users' sadness. 
\end{itemize}

This observational study presents multiple relevant findings. 
We found that users who receive toxic replies are more likely to express all three emotions compared to users who do not receive toxic replies. Furthermore, our results indicate that the amount of toxicity does not play a significant role in changing the anger or anxiety of users who already received at least one toxic reply, while higher toxicity leads to users being more likely to express more sadness in toxic conversations. Moreover, expressing emotions before the first toxic reply is likely to lead to boosting such emotions in the rest of the conversation. Finally, conversation topics are important factors that contribute to the emotional structure of the conversation. To the best of our knowledge, this is the first large-scale data-driven study that examined the impact of toxicity on users' emotions on social media, with a particular focus on Twitter. The findings in our study can help to develop prediction models of possible emotions, which can be used to provide interventions to mitigate the negative emotional impacts.

\section{Related Work}

There are different types of antisocial online behaviors, such as toxicity~\cite{xenos2022toxicity}, racist attacks against minorities~\cite{tahmasbi2021go,fredericks2021waiting,matamoros2017platformed,he2021racism}, misogynistic hatred~\cite{mantilla2013gendertrolling,parent2019social,jones2020sluts}, toxic masculinity~\cite{southern2019othering}, and others. Even though many recent studies detect antisocial behavior after such behavior occurred~\cite{chandrasekharan2017bag,kumar2017antisocial} there are some studies using certain features to predict whether the conversation will be developed in an antisocial manner~\cite{saveski2021structure,zhang2018characterizing,bao2021conversations,10.1145/3342220.3344933}.
In this study, the focus is on the literature examining the psychological and emotional impact of antisocial behavior on the victims. 


\textbf{The Psychological Consequences of Online Harassment}. 
Previous literature shows that cyberbullying leaves adverse consequences on mental health, particularly for adolescents~\cite{halliday2023relationship}. Moreover, victims are likely to commit self-harm and suicidal attempts~\cite {yang2021consequences,hinduja2010bullying,bannink2014cyber,hinduja2010bullying,messias2014school,van2014relationship,reed2015testing} as well as suffer from psychological distress~\cite{oksanen2020cyberbullying,jenaro2018systematic,brack2014cyberbullying,MARTINEZMONTEAGUDO2020112856,Alhajji,ijerph17010045,wang2019common}, depression, anxiety, and lower self-esteem~\cite{EYUBOGLU2021113730,reed2015testing,stevens2021cyber}. Also, females tend to report a higher prevalence of cyberbullying assaults and they are more likely to report distress and suicidal ideation compared to males~\cite{kim2019sex}. Other literature found a correlation between cyberbullying victimization and substance use~\cite{litwiller2013cyber,reed2015testing} while victims of online harassment might also respond by acceptance and self-blame~\cite{doi:10.1177/1461444818781324,mandau2021snaps}. 

\textbf{The Emotional Consequences of Cyberbullying}. 
Prior works showed that emotional harm is one of the victims' common experiences after online abuse and cyberbullying~\cite{sambasivan2019they,alhujailli2020affective,nixon2014current}. 
Moreover, literature found that problems with emotion regulation increase the likelihood of individuals cyberbullying others or becoming the victim of cyberbullying~\cite{baroncelli2014unique,den2016can,arato2022risk}. Furthermore, perpetrators and victims show different sets of emotions, where victims are likely to express passive emotions such as sadness, humiliation, and embarrassment~\cite{gianesini2015cyberbullying}. Also, the emotion of anger received attention from researchers investigating cyberbullying and cybervictimisation~\cite{ak2015cybervictimization}, where anger has been shown as the most common reaction to cyberbullying~\cite{campbell2012victims,beran2005cyber,ortega2012emotional} as well as sadness~\cite{raskauskas2007involvement}.

However, there are no prior data-driven studies that aimed to evaluate the impact of online attacks on individual emotions in the online setting. To the best of our knowledge, this is the first observational study analyzing social media data to examine the effect of toxicity on users' feelings of anger, anxiety, and sadness. The goal of the study is to analyze how users' emotions change after a toxic attack occurs, which could potentially lead to developing strategies to help users mitigate emotional reactivity~\cite{zhao2022cyberbullying}.

%

\section{Data Collection}

\begin{figure*}[t]
    \centerline{\includegraphics[width=0.90\textwidth]{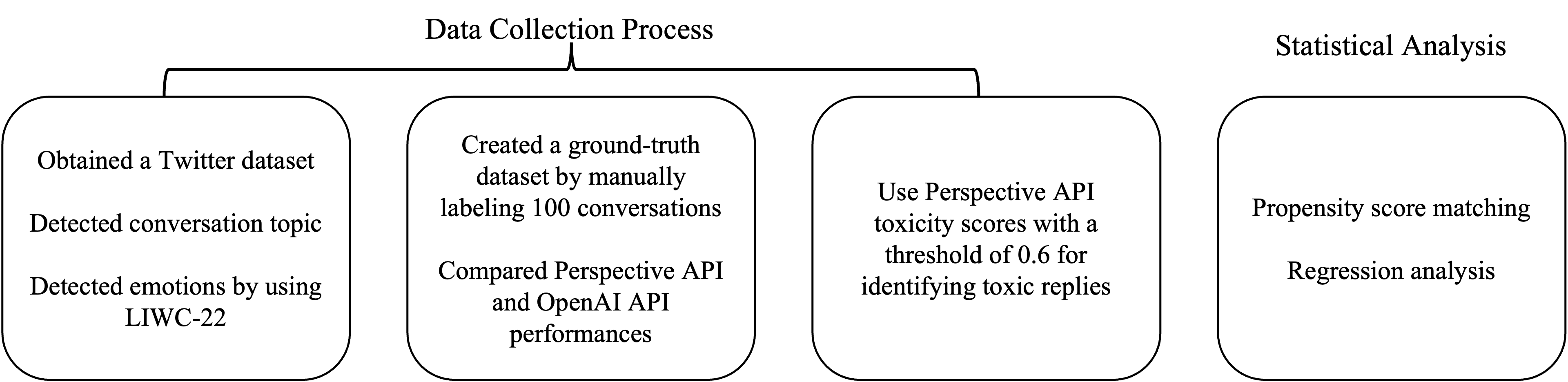}}
    \caption{Study framework.}
    \label{fig:framework}
\end{figure*}

The study framework has been shown in Figure~\ref{fig:framework}. It is demonstrated that the data collection process involved multiple steps, which will be described in more detail in this section. Firstly, we collected a Twitter dataset and detected topics and emotions, followed by the process of creating the ground-truth dataset to choose the tool to detect toxic replies with the highest accuracy. 

\textbf{Dataset}: 
The dataset analyzed originates from a recent study investigating users' reactions in online toxic conversations~\cite{aleksandric2022twitter}.
The data has been collected by utilizing Twitter API~\cite{twitterapi} to obtain a daily random sample of tweets from Aug 14 - Sep 28, 2021. The dataset includes both main tweets and their replies, as well as their toxicity scores obtained by Google's Perspective API~\cite{jigsaw2021perspective}.

\textbf{Reply Trees}: 
Similarly as in previous studies~\cite{aleksandric2022twitter,saveski2021structure},
conversations were represented as \emph{reply trees}, where a tweet is a \emph{child node} of another tweet when it is a reply to that tweet. The \emph{root} of a reply tree is the initial tweet that receives replies. The authors of the root tweets are named as \emph{root authors}. \emph{Direct replies} are located in the first layer of reply trees (replies to the root tweet) while \emph{nested replies} are located in other layers of reply trees other than the first layer, i.e., replies to replies. 
Each \emph{reply tree} has the following traits: 
\emph{Depth} referring to the depth of the conversation's deepest node (the longest path from the root tweet to last reply); \emph{Width} referring to the maximum number of tweets at any tree level. \shirin{you can add some version of that figure we had.} \ana{i will see if we have space at the end}


\textbf{Discovering Conversation Topics}: The dataset used in this study contains a random sample of Twitter conversations which can potentially include a large number of topics. However, there might be some topics that provoke more emotional responses from the users involved in the discussion. For example, users might get more angry if they receive a toxic reply concerning their political views, or they might get sad if the toxicity is directed at their health-related decisions. 
\shirin{also discuss about if and how this model outperform other models.} 
Thus, a topic classification model~\cite{antypas-etal-2022-twitter} used in previous studies~\cite{leiter2023chatgpt,hewitt2023backpack,cho2023deep,towle-zhou-2023-model} was utilized to determine the main topic of the conversation by passing the text of the main tweet as the input. Note that this model has been fine-tuned for multi-label classification on 11,267 tweets yielding 19 discussion topics such as \emph{news \& social concern}, \emph{diaries \& daily life}, \emph{business \& entrepreneurs}, and others. Scores obtained per topic are in range from 0 to 1, where a  higher score suggests that the text is more related to that topic. 

\textbf{Detecting Emotions}: 
We used LIWC-22~\cite{boyd2022development} to detect emotions of each tweet in the dataset. This tool analyzes text to provide insights into the person's emotions, social and cognitive processes, etc. It has been widely used for psychological analysis of users online~\cite{del2014study,lyu2023detecting,mukta2022predicting} and it has shown a decent performance for detecting emotions in verbal expression~\cite{kahn2007measuring}. It treats each tweet individually and provides scores for each post in range 0$-$99, representing a percentage of words in the text related to a specific attribute. 

\textbf{Creating the Ground-truth Dataset}: The next step of data collection involved detecting toxic replies in the dataset.
We first obtained tweets and their corresponding toxicity scores from the previous study~\cite{aleksandric2022twitter}. However, it was not clear which threshold should be used to detect toxic replies in the dataset. To evaluate the scores, we 
manually extracted a random sample of 50 toxic conversations that contains at least one reply with the \emph{Severe toxicity} score higher than 0.5, and 50 with no reply with the \emph{Severe toxicity} score higher than 0.5. 
The total number of tweets included in the random sample was 943 (843 replies). Then, four annotators manually labeled each tweet as 1 for toxic, and 0 if the reply is not toxic by looking at the whole conversation, trying to capture the context of conversations. Two labelers annotated 50 (25 toxic and 25 non-toxic) conversations, and the other two labelers annotated the rest. The computed Cohen's Kappa score~\cite{kvaalseth1989note} was 0.5 showing a 93.5\% agreement. Finally, the ground-truth dataset consisted of 64 (6.8\%) toxic tweets belonging to 26 conversations and 879 (93.2\%) non-toxic tweets. 

\textbf{Detecting Toxic Replies}:
As toxicity detection is a popular topic in natural language processing (NLP) literature, there are many tools that researchers developed to accomplish the task with high accuracy. Recently, ChatGPT became a tool used for different NLP tasks, such as detecting offensive language and hate speech~\cite{huang2023chatgpt,li2023hot}, stance detection~\cite{zhang2023investigating}, detoxification~\cite{tang2023detoxify}, etc. On the other hand,
Google's Perspective API has been widely used in the previous literature for toxicity detection on social media~\cite{obadimu2019identifying,ali2021understanding,chong2022understanding}. This API processes the given text input and provides output scores in the range from 0 to 1 such as \emph{Severe toxicity}, \emph{Toxicity}, \emph{Profanity}, \emph{Sexually explicit}, and others where a score closer to 1 means a higher severity for a specific attribute.
The evaluation of Perspective API received a lot of attention in previous studies. There are some works classifying texts with a score greater than 0.5 as toxic~\cite{habib2022exploring,obadimu2021developing,trujillo2022make} and others using a stricter threshold of 0.8~\cite{horta2021platform} while it is also possible to use both~\cite{shen2022xing}. However,
it is not clear whether the proposed approaches performed better than the newer tools that emerged in the meantime, such as ChatGPT. Thus, we conducted an experiment to compare the performance of Google's Perspective API and OpenAI API in the toxicity detection task. 

\textbf{Obtaining New Toxicity Scores}: Even though toxicity scores have been provided by the previous study~\cite{aleksandric2022twitter}, we still re-ran the Perspective API on the dataset as the new version of the API has been released~\cite{lees2022new}. The goal was to find a threshold that reaches the highest accuracy compared to the manually labeled sample. For that, we used \emph{Severe toxicity} and \emph{Toxicity} attributes. As the scores provided by the API are in range from 0 to 1, we increased the testing value for 0.1 in each iteration to find the most accurate threshold for classification.

\textbf{GPT Labels}: We used the OpenAI API  \emph{gpt$-$3.5$-$turbo$-$16k} version to obtain the binary toxicity labels. We passed the following prompt \emph{"Given the following post, determine if it is contextually toxic. Respond in an array format [toxic\_value, explanation], where the first element is either '1' for yes or '0' for no, and the second element is the explanation for the value."} to the API with each reply individually. However, there are two different parameters that could be changed when passing prompts to the OpenAI API: \emph{temperature} and \emph{top\_p}. 
According to endpoint documentation, it is not suggested to change both parameters at the same time~\cite{openaidoc}. Thus, as the \emph{temperature} has been defined as the randomness of the output, we decided to test how different temperatures affect the accuracy of the classification. Once again, we increased the temperature by 0.1 in each iteration to determine which temperature provided the best results. Obtained labels were compared with the manually labeled sample.

\begin{table}[h!] \centering
\caption{Comparing the accuracy of each model and corresponding thresholds.} 
\label{accuracy} 
\resizebox{0.95\columnwidth}{!}{%
\begin{tabular}{ccc|cc}
 & \multicolumn{2}{c}{{Perspective API} } & \multicolumn{2}{c}{{Open API} } \\
   Score Threshold     & Severe Toxicity   & Toxicity  & Temperature & gpt$-$3.5$-$turbo$-$16k \\
\hline
0.1 & 0.92 & 0.58 & 0.1 & 0.89\\
0.2 &  0.94& 0.74& 0.2 & 0.89\\
0.3 & 0.94 & 0.84 & 0.3 & 0.89\\
0.4 & 0.94 & 0.89& 0.4 & 0.89\\
0.5& 0.93 & 0.93& 0.5 & 0.89\\
0.6&  0.93& 0.95& 0.6 & 0.88\\
0.7& 0.93 & 0.94& 0.7 & 0.88\\
0.8& 0.93 & 0.94& 0.8 & 0.88\\
0.9& 0.93 & 0.93 & 0.9 & 0.87\\
\hline
\end{tabular}
}
\end{table}

\textbf{Evaluating the performance}: Finally, we were able to compare the labels obtained from Perspective API and OpenAI to the manually labeled sample.
Based on the results presented in Table~\ref{accuracy}, we found that the Toxicity attribute outperforms both Severe Toxicity and gpt$-$3.5$-$turbo$-$16k with the highest accuracy of 0.95 (95\%) for the threshold of 0.6. Therefore, we will use this threshold to detect toxic replies in the dataset. Any reply that shows a score equal to or higher than 0.6 in the Toxicity is considered as \emph{toxic}.


In the dataset, there are 20,544 (3.9\%) toxic tweets belonging to 13,172 (16.5\%) conversations, where 11,784 (57.4\%) of the toxic tweets were posted by users other than the root author. On the other hand, there are 507,497 tweets belonging to 79,580 conversations that are not considered toxic.

\section{Independent, Dependent, and Control Variables}
\label{variables}
Causal inference is used paired with multivariate regression analysis to find causal associations between different levels of toxicity and users' emotions. In more detail, we examine how toxicity impacts users' emotions and what other factors contribute to triggering anger, anxiety, and sadness in Twitter users. 

\textbf{Dependent Variables} 
\label{dep_vars}
included in the analysis are users' emotions where we describe two sets of variables, the average emotions of root authors in whole conversations, and the average emotions of root authors after the first toxic reply occurs in a toxic conversation. The first set of variables was included as the goal is to compare the emotions expressed throughout the conversation by root authors who received toxic replies and root authors who did not. The second set aims to clarify how emotions change after receiving the first toxic reply and whether the amount of toxicity plays a significant role in emotional reactions.
The following variables were used in the models: 
(1)~\emph{anxiety}: a numeric variable representing the average root author's anxiety in a conversation.
(2)~\emph{anger}: a numeric variable representing the average root author's anger in a conversation.
(3)~\emph{sadness}: a numeric variable representing the average root author's sadness in a conversation.
(4)~\emph{anxiety\_after}: a numeric variable representing the average root author's anxiety after a toxic reply occurs in a conversation.
(5)~\emph{anger\_after}: a numeric variable representing the average root author's anger after a toxic reply occurs in conversation.
(6)~\emph{sadness\_after}: a numeric variable representing the average root author's sadness after a toxic reply occurs in a conversation.
Note that all the dependent variables were rounded up to the closest integer for the analysis purposes. In addition, we are able to calculate the emotions after the toxic reply only in the conversations containing toxic replies.

\textbf{Independent Variables} 
consist of computed percentages of toxic replies in the conversations. 
Note that we consider the location of the toxicity as well as the level of toxicity in the conversation reply tree. Thus, we considered the following independent variables: 
(1)~\emph{direct\_toxicity}, a numeric variable representing a ratio of the toxic direct replies to the total number of direct replies in the conversation.  
(2)~\emph{nested\_toxicity}, a numeric variable representing a ratio of the number of toxic nested replies to the total number of nested replies in the conversation. \shirin{it's good to have a figure to refer to it. People can get confused by these definitions of directed and nested.}

\textbf{Control Variables}
While evaluating the effect of toxicity on Twitter users' emotions, we need to include certain variables as confounding factors. 
In more detail, we included features related to conversation structure, the emotions of the root tweet, the topic of the root tweet, and users' \emph{activity}, \emph{visibility}, and \emph{identifiability}.  
\textbf{Users' online activity} contains \emph{num\_friends}, \emph{num\_tweets} and \emph{account\_age} (in years) as numeric variables. Such variables can affect how users emotionally respond to toxic content. For example, users who post a lot might not be so emotionally affected by toxic replies, while users whose accounts are younger might care more about their reputation and show emotions of anger, sadness, or anxiety more compared to older accounts.
\textbf{Online visibility} contains  \emph{num\_followers}, and \emph{listed\_counts} as numeric variables and \emph{verified} as a binary variable.  
Previous literature ~\cite{elsherief2018peer} found that there is a relationship between online visibility and receiving hate. Therefore, there might be a correlation between online visibility and how users emotionally respond to that hate. For example, verified accounts might not emotionally react to toxic content, but accounts that have fewer followers might express anger, anxiety, or sadness more in these situations.
\textbf{Identifiability} consists of profile characteristics that help identify a user, such as profile \emph{description\_length} (in characters) as a numeric variable, and \emph{has\_URL}  and \emph{has\_location} as binary variables, indicating whether the profile contains URLs to other user-related websites and whether the user provided the location on their profile. A variable \emph{has\_image} was also collected, but all users in our dataset had images provided. Previous literature suggested that anonymous accounts show more abusive behavior compared to other identifiable accounts~\cite{schlesinger2017situated,correa2015many,zhang2014anonymity}. Therefore, we believe that in this study, it is possible that anonymous accounts might show less severe emotions of anger, anxiety, and sadness when receiving toxic content compared to more identifiable users.
\textbf{Emotions before a toxic reply}: furthermore, we believe that if the user was already expressing a certain emotion before the toxic reply, it is important to acknowledge whether their emotions changed or not. For example, if the user is already angry and receives a toxic comment, the user would either become more angry or their emotions might remain the same. Therefore, we included the following control variables in the analysis to find the impact of the toxicity on users' emotions: (1) \emph{anger\_before}: a numeric variable representing the average root author's anger before a toxic reply occurs in conversation; (2) ~\emph{anxiety\_before} representing a numeric variable indicating the average root author's anxiety before a toxic reply occurs in a conversation, and (3) ~\emph{sadness\_before}: a numeric variable representing the average root author's sadness before a toxic reply occurs in a conversation. Once again, such variables are only computed for conversations with toxic replies. Also, they were computed by chronologically ordering the tweets within the conversation and calculating averages before the toxic reply occurred.
\textbf{Conversation structure}  
consists of \emph{width} and \emph{depth} as numeric variables. We believe that the users' emotions might be affected depending on how big the conversation is. For example, users might not feel the same if they received a couple of toxic comments in a very large online discussion, while they might express emotions more in smaller conversations.
We also included \textbf{root\_toxicity} indicating if the root tweet is toxic or not. Root authors that share already toxic tweets might receive more toxicity while they might express emotions of anger more than others. 
Finally, all 19 conversation topic variables were also included as confounding factors as there might be a case that the main topic of the conversation affects how users emotionally respond to toxic content. For example, political or daily life topics might attract more emotional responses compared to other topics.

\section{Descriptive Statistics of Variables}

\begin{table*}[t!] \centering 
\caption{Characteristics of conversations in our dataset.}
\resizebox{0.60\textwidth}{!}{%
\begin{tabular}{lllll|llll}
\hline
\hline
& \multicolumn{4}{c}{Conversations with Toxic Replies} & \multicolumn{4}{c}{Conversations without Toxic Replies} \\
\hline
 Characteristics & Min    & Median & Mean     & Max  & Min    & Median & Mean     & Max \\
\hline
anxiety &0 & 0 & 0.15 & 50 & 0  &  0 &  0.13 & 100  \\
sadness       & 0  &  0 & 0.48 & 40  &0   &  0 &  0.52  &  100 \\
anger          & 0 &  0 & 0.28 & 50 & 0   &  0 &  0.18 & 100   \\
anxiety\_before  &0 & 0 & 0.15 & 50 & NA  &  NA &  NA & NA  \\
sadness\_before      & 0  &  0 & 0.48 & 40  &NA  &  NA &  NA & NA \\
anger\_before          & 0 &  0 & 0.28 & 50 & NA  &  NA &  NA & NA  \\
anxiety\_after  &0 & 0 & 0.15 & 50 & NA  &  NA &  NA & NA  \\
sadness\_after      & 0  &  0 & 0.48 & 40  &NA  &  NA &  NA & NA \\
anger\_after          & 0 &  0 & 0.28 & 50 & NA  &  NA &  NA & NA   \\
\hline
direct\_toxicity &0 & 0.25& 0.4& 1 & NA  &  NA &  NA & NA   \\
nested\_toxicity& 0 & 0& 0.08 & 1  & NA  &  NA &  NA & NA   \\
\hline
arts \& culture  & 0.001 & 0.02 & 0.05 & 0.88 & 0.001 & 0.03 & 0.06 & 0.92  \\
business \& entrepreneurs  & 0.001 & 0.01 & 0.04 & 0.97  & 0 & 0.01 & 0.05 & 0.98 
\\
celebrity \& pop culture   & 0.003  & 0.03 & 0.12 & 0.99
 &0 & 0.03& 0.12& 0.99\\
diaries \& daily life  & 0.005 & 0.29 & 0.37 & 0.98 
 &0 & 0.3& 0.38 & 0.98 \\
family   & 0.001 & 0.01 & 0.03 & 0.92
 & 0 & 0.01 & 0.04 & 0.95\\
fashion \& style   &0.0005 & 0.005 & 0.03 & 0.98 
 & 0  & 0.01 & 0.03 & 0.98\\
film tv \& video  & 0.002 & 0.02 & 0.1 & 0.99 
 & 0 & 0.02& 0.1 & 0.99\\
fitness \& health  &0.001 & 0.01 & 0.03 & 0.98 
 & 0 & 0.01 & 0.03 & 0.98\\
food \& dining  & 0.0004 & 0.004 & 0.04 & 0.97 
 &0 & 0.004& 0.04& 0.97 \\
gaming  &0.001 & 0.01 & 0.04 & 0.95
 &0 &0.01& 0.03& 0.96 \\
learning \& educational  & 0.001 & 0.01 & 0.03 & 0.92
 & 0 & 0.01 & 0.04 & 0.94 \\
music  & 0.001 & 0.01 & 0.07 & 0.99 
 & 0  & 0.01& 0.08& 0.99 \\
news \& social concern  & 0.001 & 0.07 & 0.25 & 0.99
 & 0& 0.05 & 0.18& 0.99\\
other hobbies & 0.003 & 0.05 & 0.09 & 0.8
& 0 & 0.05& 0.1 & 0.84\\
relationships &0.001 & 0.02 & 0.07 & 0.9 
 &0 & 0.02& 0.09& 0.93\\
science \& technology  &0.001 & 0.01 & 0.03 & 0.96
 &0 & 0.01& 0.03 & 0.96\\
sports  & 0.0004 & 0.01 & 0.13 & 0.99
 & 0  & 0.01 & 0.13 & 0.99\\
travel \& adventure & 0.001 & 0.01 & 0.02 & 0.91
 &0 & 0.01& 0.03& 0.93 \\
youth \& student life  & 0.001 & 0.005 & 0.02& 0.91
 &0  & 0.005 & 0.02& 0.91 \\
\hline
 \# conversations   & \multicolumn{4}{c}{7,205}        & \multicolumn{4}{c}{72,594}  \\
\hline
\end{tabular}
\label{stats} 
}
\end{table*}

\begin{figure*}[t]
    \centerline{\includegraphics[width=0.95\textwidth]{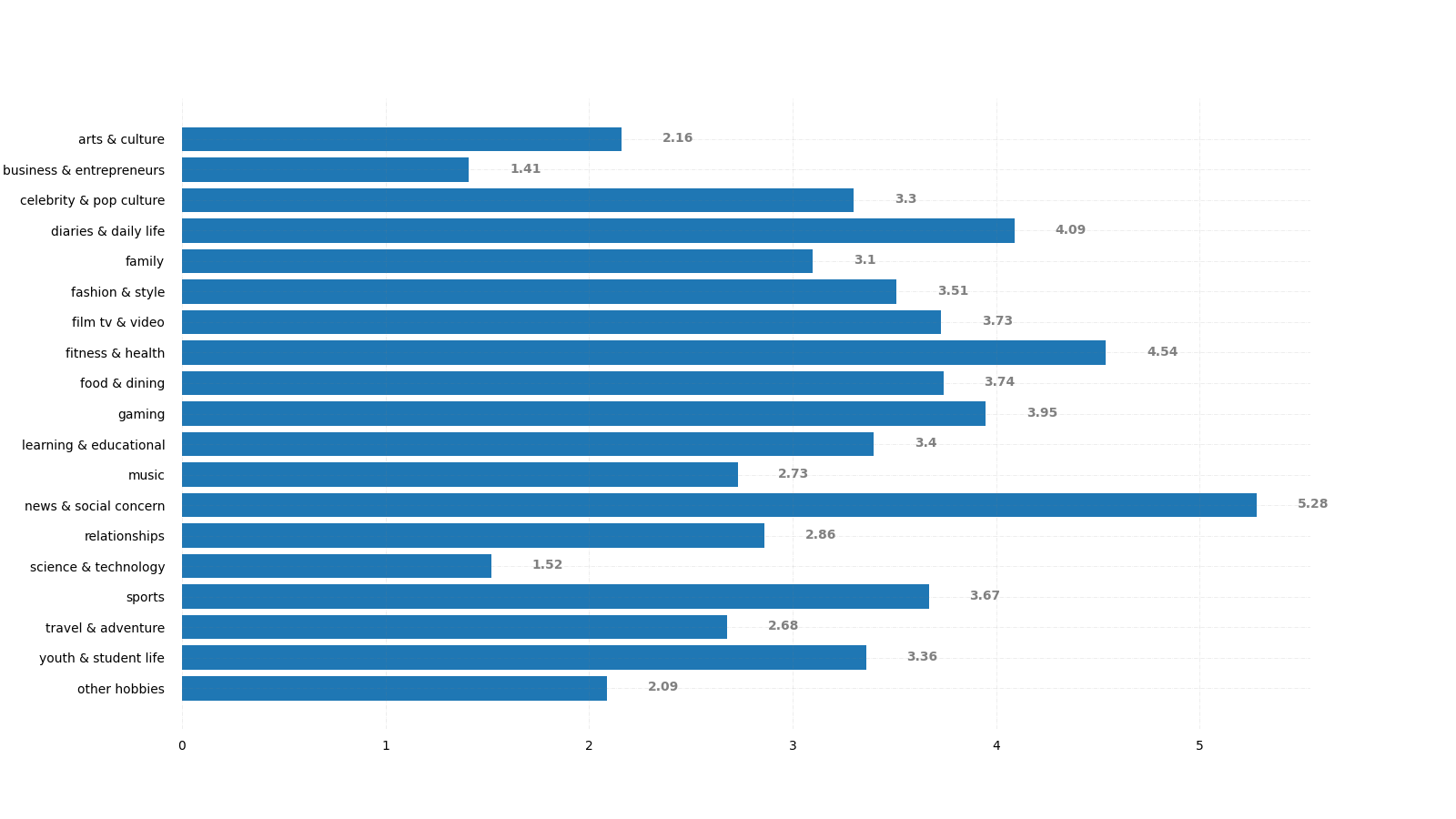}}
    \caption{Percentage of replies being toxic per topic.}
    \label{fig:topics}
\end{figure*}

This section provides detailed descriptive statistics on the emotions and topics expressed in the conversations.
We compare the prevalence of emotions and topics discussed in conversations with and without toxic replies. Table~\ref{stats} includes minimum, median, mean, and maximum values for topics and average emotions of the root authors in conversation with and without toxic replies. Interestingly, we observed that the mean anxiety and anger were higher in conversations with toxic replies being 0.15 vs. 0.13 and 0.28 vs. 0.18, respectively. On the other hand, the mean sadness expressed by the root authors is higher in conversations without toxic replies (0.52 vs 0.48), which can potentially indicate that users who receive toxic replies emotionally respond in an anxious or angry manner. Note that the mean direct\_toxicity is significantly higher compared to nested\_toxicity (0.4 vs. 0.08).
Furthermore, the most prevalent topics discussed in our conversations are \emph{diaries \& daily life}, \emph{news \& social concern}, and \emph{sports} in both datasets, suggesting that these topics are most widely discussed on the platform in general. 

We believe that certain topics from the list might naturally trigger more toxicity compared to others. Thus, we created a plot in Figure~\ref{fig:topics}, to discover the relationship between the percentage of replies being toxic and conversation topics. The topics that lured the most toxic replies were news \& social concern, fitness \& health, and diaries \& daily life.
On the other hand, topics that received the least toxic replies were business \& entrepreneurs and science \& technology, receiving only 1.41\% and 1.52\% of replies that are toxic, respectively. This might indicate that social media users tend to attack others based on their opinions about social and everyday life, while discussions about professional development do not lead to conversations developing in a toxic manner. 
\shirin{if space is needed, we can remove either the figure or the stats on topics from the table. one is enough.}

\section{Analysis and Results}

This section describes the statistical analysis performed to test the formulated hypotheses. The first step of analysis involved performing propensity score matching, a technique employed for balancing two datasets~\cite{yao2017reporting} so the conversation dataset without toxic replies includes the same number of conversations as the dataset with toxic replies. In more detail, the main purpose of propensity score matching is balancing treatment and control groups by choosing the observations with a similar propensity of receiving treatment  in order to provide insights about the causal impact of the treatment~\cite{10.2307/2288398} (in our case, receiving toxicity). The scores were calculated by using the user's characteristics as independent variables in the logistic regression model where the dependent variable was a binary variable indicating whether the conversation received toxic replies or not. Therefore, the balanced dataset included the same number of conversations with and without toxic replies (7,205), suggesting that the whole dataset included 14,410 conversations in total. 

The first part of the analysis included three multivariate Poisson regression models where we investigated causal associations between root authors' emotions and receiving toxicity (testing H1-3). Therefore, the dependent variables in these models were root authors' average emotions expressed in conversations (anger, anxiety, and sadness), while independent variables were direct\_toxicity and nested\_toxicity. This analysis was performed on the balanced dataset described above, in order to compare the emotions of root authors of conversations with and without toxic replies. The Poisson model was the most suitable for our dataset, as all the dependent variables were count variables very skewed to the right.

The second part only involved the analysis of toxic conversations, where conversations that did not include root authors' comments after the first toxic replies were discarded. Therefore, the number of conversations with toxic replies in this analysis was 3,408 conversations. The goal of this analysis is to find out how the amount of toxicity affects the emotions of users who received toxic replies, and whether their emotions expressed before the toxic attack occurred play a significant role in their emotional responses (testing H4-H6). The dependent variable of the multivariate Poisson regression models were average emotions after the first toxic comment took place (anxiety\_after, anger\_after, and sadness\_after), where we also included average emotions before the first toxic replies as control variables besides other variables discussed in section~\ref{variables}. 

Furthermore, to address possible bias due to multiple hypotheses testing, we use Bonferroni correction~\cite{armstrong2014use}. Therefore, we divided the $p-value$ of 0.05 by the total number of hypotheses, yielding a value of 0.008. Thus, the $p-values$ lower than 0.008 would signify statistical significance in the analysis. Finally, even though propensity score matching has been used in the literature to isolate the effect of the treatment, we cannot certainly claim that we are able to identify causal inferences without having any other confounding factors that also impact users' emotions. Therefore, the results presented are not causations; they represent causal associations. 

\begin{table*}[!htbp] \centering 
  \caption{Regression Models 
  } 
  \label{results} 
  \resizebox{0.95\textwidth}{!}{%

\begin{tabular}{@{\extracolsep{5pt}}lcccccc} 
\\[-1.8ex]\hline 
\hline \\[-1.8ex] 
 & \multicolumn{6}{c}{\textit{Dependent variable:}} \\ 
\cline{2-7} 
\\[-1.8ex] & anxiety & anger & sadness & anxiety\_after & anger\_after & sadness\_after \\ 
\\[-1.8ex] & M1 & M2 & M3 & M4 & M5 & M6\\ 
\hline \\[-1.8ex] 
 direct\_toxicity& 0.227$^{**}$ (0.064) & 0.576$^{***}$ (0.047) & 0.114$^{**}$ (0.033) & $-$0.096 (0.159) & $-$0.001 (0.102) & 0.535$^{***}$ (0.056) \\ 
  nested\_toxicity & $-$1.136$^{***}$ (0.254) & 0.702$^{***}$ (0.121) & $-$0.126 (0.102) & $-$1.613$^{*}$ (0.568) & 0.685 (0.268) & $-$0.450 (0.177) \\ 
  width & $-$0.006$^{*}$ (0.002) & $-$0.006$^{**}$ (0.002) & $-$0.016$^{***}$ (0.002) & $-$0.012 (0.006) & $-$0.025$^{***}$ (0.006) & $-$0.002 (0.002) \\ 
  depth & $-$0.004 (0.006) & 0.001 (0.004) & 0.003 (0.003) & 0.004 (0.006) & $-$0.006 (0.007) & $-$0.012 (0.005) \\ 
  root\_toxicity True & $-$0.283$^{**}$ (0.083) & 0.545$^{***}$ (0.048) & $-$0.090 (0.039) & $-$1.563$^{***}$ (0.265) & 0.427$^{***}$ (0.084) & 0.249$^{***}$ (0.051) \\ 
  num\_followers & $-$0.0 (0.0) & $-$0.0 (0.0) & $-$0.0$^{***}$ (0.0) & 0.0 (0.0) & 0.0 (0.0) & $-$0.0 (0.0) \\ 
  num\_friends & $-$0.0 (0.0) & $-$0.00002$^{*}$ (0.00001) & $-$0.0 (0.0) & 0.00001 (0.00001) & $-$0.0001$^{**}$ (0.00003) & $-$0.00004$^{*}$ (0.00001) \\ 
  num\_tweets & $-$0.0 (0.0) & $-$0.0 (0.0) & 0.0$^{***}$ (0.0) & 0.0 (0.0) & 0.0 (0.0) & 0.0$^{***}$ (0.0) \\ 
  listed\_counts & $-$0.0 (0.00003) & 0.00001 (0.00002) & 0.0001$^{***}$ (0.00001) & $-$0.001 (0.001) & 0.0004 (0.0002) & $-$0.001 (0.0005) \\ 
  description\_length & 0.002$^{***}$ (0.0005) & $-$0.0002 (0.0004) & $-$0.006$^{***}$ (0.0003) & $-$0.001 (0.001) & 0.003$^{**}$ (0.001) & $-$0.005$^{***}$ (0.0004) \\ 
  verified True & $-$0.223 (0.113) & 0.402$^{***}$ (0.081) & $-$0.539$^{***}$ (0.085) & 0.169 (0.518) & $-$0.528 (0.314) & 0.010 (0.254) \\ 
  account\_age & 0.002 (0.006) & $-$0.006 (0.005) & $-$0.034$^{***}$ (0.003) & 0.001 (0.014) & $-$0.004 (0.010) & $-$0.045$^{***}$ (0.006) \\ 
  has\_location True & 0.195$^{**}$ (0.057) & 0.158$^{**}$ (0.044) & 0.307$^{***}$ (0.030) & 0.889$^{***}$ (0.163) & 0.307$^{**}$ (0.089) & 0.679$^{***}$ (0.058) \\ 
  has\_url True & $-$0.100 (0.048) & $-$0.136$^{**}$ (0.039) & $-$0.023 (0.025) & $-$0.363$^{*}$ (0.113) & $-$0.156 (0.073) & 0.108 (0.041) \\ 
  anxiety\_before &  &  &  & 0.140$^{***}$ (0.013) &  &  \\ 
  anger\_before &  &  &  &  & 0.063$^{***}$ (0.007) &  \\ 
  sadness\_before &  &  &  &  &  & 0.102$^{***}$ (0.004) \\ 
  arts\_.\_culture & $-$0.571 (0.311) & $-$0.413 (0.260) & $-$0.465$^{*}$ (0.164) & $-$0.275 (0.720) & 2.112$^{***}$ (0.326) & $-$0.854$^{*}$ (0.270) \\ 
  business\_.\_entrepreneurs & $-$0.681$^{*}$ (0.249) & $-$1.406$^{***}$ (0.252) & $-$1.340$^{***}$ (0.186) & 2.012$^{***}$ (0.364) & $-$3.035$^{***}$ (0.687) & $-$4.614$^{***}$ (0.624) \\ 
  celebrity\_.\_pop\_culture & $-$0.613$^{***}$ (0.143) & $-$0.405$^{**}$ (0.121) & 0.090 (0.065) & 0.173 (0.349) & 0.208 (0.211) & 0.168 (0.106) \\ 
  diaries\_.\_daily\_life & 0.517$^{***}$ (0.122) & 0.668$^{***}$ (0.097) & 0.554$^{***}$ (0.065) & $-$0.211 (0.276) & 0.338 (0.188) & 0.122 (0.111) \\ 
  family & 0.216 (0.335) & $-$0.219 (0.272) & 1.271$^{***}$ (0.146) & 0.271 (0.605) & $-$1.685$^{*}$ (0.516) & $-$0.058 (0.268) \\ 
  fashion\_.\_style & 0.198 (0.206) & $-$0.384 (0.217) & $-$0.310 (0.127) & $-$3.169 (1.208) & $-$0.620 (0.467) & 1.084$^{***}$ (0.181) \\ 
  film\_tv\_.\_video & 0.412$^{***}$ (0.105) & 0.010 (0.096) & 0.316$^{***}$ (0.056) & $-$1.585$^{***}$ (0.387) & $-$1.280$^{***}$ (0.236) & 0.349$^{**}$ (0.095) \\ 
  fitness\_.\_health & 0.390 (0.158) & $-$0.804$^{***}$ (0.194) & $-$0.276 (0.139) & 0.777 (0.361) & $-$0.973$^{*}$ (0.364) & $-$1.691$^{***}$ (0.373) \\ 
  food\_.\_dining & $-$3.546$^{***}$ (0.508) & $-$0.597$^{***}$ (0.151) & $-$0.758$^{***}$ (0.098) & $-$3.005$^{*}$ (0.922) & $-$0.574 (0.281) & 0.258 (0.122) \\ 
  gaming & $-$1.003$^{***}$ (0.232) & 0.433$^{**}$ (0.116) & $-$0.860$^{***}$ (0.123) & 0.923$^{*}$ (0.284) & $-$0.208 (0.256) & $-$1.549$^{***}$ (0.252) \\ 
  learning\_.\_educational & $-$0.245 (0.582) & 0.912 (0.481) & $-$0.017 (0.350) & $-$0.552 (1.394) & $-$1.826 (0.859) & 2.711$^{***}$ (0.478) \\ 
  music & $-$0.262 (0.145) & $-$0.505$^{**}$ (0.135) & 0.006 (0.067) & $-$0.670 (0.401) & $-$0.923$^{**}$ (0.269) & 0.179 (0.116) \\ 
  news\_.\_social\_concern & $-$0.032 (0.093) & $-$0.040 (0.078) & $-$0.637$^{***}$ (0.058) & 0.186 (0.210) & 0.070 (0.145) & $-$0.590$^{***}$ (0.102) \\ 
  other\_hobbies & $-$2.768$^{***}$ (0.326) & $-$1.743$^{***}$ (0.235) & $-$1.494$^{***}$ (0.150) & $-$0.740 (0.598) & $-$2.330$^{***}$ (0.467) & $-$0.317 (0.240) \\ 
  relationships & $-$1.412$^{***}$ (0.262) & $-$0.858$^{***}$ (0.198) & $-$1.614$^{***}$ (0.127) & 0.022 (0.507) & 0.537 (0.335) & $-$0.098 (0.189) \\ 
  science\_.\_technology & $-$0.704 (0.293) & $-$0.054 (0.229) & $-$1.380$^{***}$ (0.240) & $-$2.342 (0.906) & 2.871$^{***}$ (0.252) & 0.891$^{*}$ (0.291) \\ 
  sports & $-$1.756$^{***}$ (0.159) & $-$0.171 (0.085) & $-$0.668$^{***}$ (0.065) & $-$1.408$^{***}$ (0.345) & $-$0.057 (0.170) & $-$0.314$^{*}$ (0.113) \\ 
  travel\_.\_adventure & $-$1.579$^{**}$ (0.408) & $-$1.742$^{***}$ (0.353) & $-$1.450$^{***}$ (0.216) & 0.657 (0.571) & $-$1.124 (0.615) & 0.073 (0.272) \\ 
  youth\_.\_student\_life & 0.678 (0.672) & $-$1.854$^{*}$ (0.621) & $-$0.252 (0.412) & 1.021 (1.622) & 1.152 (1.050) & $-$4.649$^{***}$ (0.727) \\ 
 \hline \\[-1.8ex] 
Observations & 14,410 & 14,410 & 14,410 & 3,408 & 3,408 & 3,408 \\ 
Log Likelihood & $-$8,006.427 & $-$11,304.250 & $-$21,059.620 & $-$1,525.933 & $-$2,971.058 & $-$6,981.386 \\ 
\hline 
\hline \\[-1.8ex] 
\textit{Note:}  & \multicolumn{6}{r}{$^{*}$p$<$0.008; $^{**}$p$<$0.001; $^{***}$p$<$1e-04} \\ 
\end{tabular} 
}
\end{table*}



\shirin{whenever we can you should make connection between our findings and the literature. }

\textbf{H1: Users receiving toxic replies are more likely to express anxiety compared to users who did not receive any toxic replies}. The results from the regression model are presented in Table~\ref{results} (M1). The model suggests that there is a positive statistically significant relationship between the direct\_toxicity and average anxiety of the root author of the conversation ($p < 0.001$). In other words, the larger percentage of replies being toxic that are direct responses to the main tweet is likely to increase the average anxiety expressed by the main user throughout the conversation. On the other hand, a relationship between nested\_toxicity and anxiety is statistically significant and negative ($p < 0.0001$), suggesting that the higher percentage of nested replies being toxic is associated with the lower anxiety of the main user. The reason can be that users might not feel personally attacked when the bigger toxic thread occurs while receiving toxic replies directly to their main posts might make them feel more anxious as they might consider such attacks more personal. Moreover, nested replies might be less visible to the root authors compared to direct replies to their tweets. 
Furthermore, root authors who started the conversation with the toxic tweet are less likely to express anxiety compared to other users ($p < 0.001$). Finally, users who provide longer profile descriptions ($p < 0.0001$) and locations ($p < 0.001$) tend to express more anxiety, indicating that more identifiable accounts might get more anxious about their content compared to more anonymous users. \emph{Therefore, the results of M1 support H1 partially, suggesting that root authors who receive more toxic direct replies are more likely to express anxiety}. Once again, it is important to note that the results obtained represent causal associations, not causations.

\textbf{H2: Users receiving toxic replies are more likely to express anger
compared to users who did not receive any toxic replies}. Model results presented in Table~\ref{results} (M2) demonstrate a positive statistically significant correlation between both direct\_toxicity and nested\_toxicity with anger ($p < 0.0001$). In more detail, root authors receiving either toxic direct or nested replies tend to express more anger. 
Moreover, root authors who initiated a conversation with the toxic tweet are likely to express more anger compared to users who started a conversation with a non-toxic tweet ($p < 0.0001$). Interestingly, the model suggests that verified accounts are more likely to convey anger compared to other users ($p < 0.0001$). It could potentially mean that verified users care about the replies shared on their posts and get angry once their point of view is under attack, or that they feel more comfortable to express their anger without worrying about its consequences.   
\emph{In summary, M2 supports our hypothesis that users receiving toxic replies are more likely to express anger compared to other Twitter users}. 

\textbf{H3: Users receiving toxic replies are more likely to express sadness compared to users who did not receive any toxic replies}. Our findings (Table~\ref{results} M2) illustrate that the association between direct\_toxicity and sadness is positive and statistically significant ($p < 0.001$), meaning that users who receive direct toxic replies on their main post are likely to get sadder compared to users who did not receive toxic replies. In addition, users who are participants of more public lists and younger accounts tend to express more sadness ($p < 0.0001$), which can potentially be due to such accounts being more concerned about the opinions of their friends and followers and therefore, expressing sadness when under attack. \shirin{it's a weird justification. how sadness and reputation related? I can understand anger and reputation better. } \ana{what about now?} \shirin{it's interesting. not sure. I think we will get some comment on this.} 
However, verified users tend to convey less sadness than others ($p < 0.0001$). As shown in M2, verified accounts tend to communicate more anger rather than sadness which can potentially indicate that such users insist on their beliefs on social media. \emph{Once again, our results partially support H3, suggesting that authors who receive toxic direct replies are more likely to express sadness compared to users who did not receive toxic replies}.

\textbf{H4: A larger amount of toxicity will likely increase users’ anxiety}. According to the results presented in Table~\ref{results} (M4), there exists a negative significant relationship between nested\_toxicity and anxiety\_after ($p < 0.008$), while the association between direct\_toxicity and anxiety\_after does not show statistical significance.
Thus, we can conclude that even though users who receive toxic replies are likely to express more anxiety compared to users who did not receive toxic replies, the amount of toxicity itself does not increase the anxiety in the first group of users, \emph{rejecting H4}. However, we observe that users who were already feeling anxious before the first toxic reply took place tend to express more anxiety after, as the correlation between anxiety\_before and anxiety\_after is positive and significant ($p < 0.0001$).

\textbf{H5: A larger amount of toxicity will likely increase users’ anger}. As demonstrated in Table~\ref{results} (M5), the relationship between toxicities and anger of users after the first toxic reply is not statistically significant ($p > 0.008$), \emph{rejecting H5}. We infer that users receiving toxic replies are likely to express more anger compared to others, however, the amount of toxicity they receive does not affect their anger. Similarly, as in testing H4, we noticed that the association between the average anger of users before the first toxic reply and average anger after the first toxic reply is positive and significant ($p < 0.0001$), meaning that users who were already angry are likely to express this emotion more after receiving toxicity. 

\textbf{H6: A larger amount of toxicity will likely increase users’ sadness}. As shown in Table~\ref{results} (M6), there is a statistically significant positive relationship between direct\_toxicity and sadness of users after the first toxic reply ($p < 0.0001$). In other words, if the user receives more toxic replies on their main posts, such users tend to express more sadness. Also, the model reveals that users showing more sadness before the first toxic reply tend to get even more sad after the first toxic reply occurs ($p < 0.0001$). Such findings signify that the amount of toxicity is likely to impact the amount of sadness users convey, \emph{partially supporting H6}. 

\textbf{Examining the impact of conversation topics on users' emotions}. Here, we compare the impact of specific topics in our models that showed as significant contributors to users' emotions in conversations. For example, in models M1, M2, and M3, we found that the conversation topic diaries\_.\_daily\_life is associated significantly with elevated anxiety, anger, and sadness, indicating that users might get emotional when discussing daily life matters, which can be more personal. On the other hand, in the same models, food\_.\_dining shows a negative correlation with all three emotions, signifying that such discussions do not trigger users' emotions. Interestingly, we found that the topic of gaming increases anger, while decreasing anxiety and sadness, meaning that the users get angry rather than anxious or sad which aligns with existing literature stating that aggression is perceived as more normal in online gaming than in offline setting~\cite{hilvert2020m}. Some of the other topics, such as relationships, science\_.\_technology, other\_hobbies, travel\_.\_adventure, and business\_.\_entrepreneurs were also found as likely to decrease all emotions discussed. Additional research is required to understand the reasoning behind such a phenomenon. 

\section{Discussion}
This study examines the emotional responses of users to toxic replies they receive on their tweets. For that, we analyzed a large dataset consisting of 79,799 conversations where 7,205 conversations contained at least one toxic reply. Then, we detected toxic replies, emotions, and topics in these conversations and performed causal association analysis by leveraging propensity score matching to balance two datasets. Our results contribute to the general understanding of the way toxicity impacts users' emotions. 
For example, we showed that users who received toxic replies on their tweets are more likely to express anger, anxiety, and sadness throughout the conversation compared to users who did not receive any toxic comments. However, problematic emotion regulation is associated with a higher likelihood of people becoming victims or cyberbullying others~\cite{baroncelli2014unique,den2016can,arato2022risk}. Furthermore, our results align with existing literature stating that victims of cyberbullying express emotions of anger~\cite{campbell2012victims,beran2005cyber,ortega2012emotional}, sadness~\cite{gianesini2015cyberbullying,raskauskas2007involvement}, and anxiety~\cite{EYUBOGLU2021113730,reed2015testing,stevens2021cyber}.
This way, we demonstrate that there is a need for more research to dive deeply into the users' reactions to toxic content and establish the framework for the implementation of advanced moderation techniques to help users emotionally cope with the toxicity. In addition, this is the first observational data-driven study investigating the way users' emotions change after receiving a toxic reply, where we showed that sadness after the first toxic reply is likely to increase as the amount of toxicity grows. Moreover, we demonstrated that users who already expressed emotions of anger, sadness, and anxiety before receiving toxic replies might be more vulnerable to toxicity attacks as their emotions are likely to increase after the toxic reply occurs. At the same time, certain findings raise concerns, as previous studies showed that not coping successfully with anger can potentially lead to further cyberbullying behavior~\cite{den2016can}. 

Despite the relevant findings presented in this study, there are obvious limitations that have to be mentioned. Firstly, tools used for detecting topics, toxic replies, and emotions are not perfect and could potentially contain biases in their classifications. 
Also, the dataset used in the study contains posts originating from unique users and, therefore, we believe that the nature of the data prevents us from analyzing the long-term emotional consequences of receiving toxicity.
Therefore, the study results indicate a causal association between receiving toxicity and expressing emotions of anger, anxiety, and sadness, however, further research is needed to prove a final causal relationship between these variables.

\section{Conclusion}
In conclusion, in this study, we formulated six hypotheses aiming to explores the emotional response of Twitter users to toxic replies they receive on their posts. 
Our preliminary findings show that receiving toxicity is likely to in certain significantly increase users' emotions of anger, anxiety, and sadness, whereas the user characteristics and conversation topic play a significant role in the ways users emotionally react to toxic comments. In summary, the presented findings provide a better understanding of the ways users' emotions change after receiving toxic replies and they represent the initial step leading to building a moderation framework that would help all social media users emotionally cope with toxic content.

\bibliographystyle{ACM-Reference-Format}
\bibliography{references}

\end{document}